WIK Wissenschaftliches Institut
für Kommunikationsdienste GmbH

# Prospects for improving competition in mobile roaming

**Ulrich Stumpf**

# Contents

## Abstract

The ability to make international roaming calls is of increasing importance to customers. However, there are various complaints that prices of retail roaming are intransparent, rigid and at levels that are unrelated to the cost of carriage. The focus if the paper is on wholesale roaming, which is the prime determinant of retail roaming prices. The paper analyses the structural conditions of wholesale roaming markets that have impaired incentives to competition, namely (1) high combined market share of the two leading operators combined with second mover disadvantages, and (2) demand externalities associated with customer ignorance and lack of control over network selection. The paper argues that a number of developments are under way that are likely to modify this situation in the future. With the introduction of SIM over-the-air programming, home mobile operators will be able to direct customers to networks with the lowest charges. As dual mode handsets become ubiquitous and as new entrant GSM 1800 operators reach nationwide coverage, second-mover disadvantages will disappear. Given the relatively small roaming volumes that GSM 1800 operators currently provide, they should have an incentive to lower charges in exchange for preferred roaming status. On the demand side of wholesale roaming markets, it will be the larger GSM 900 operators, and in particular those with a pan-European footprint, that will ask for lower charges in exchange for preferred roaming status. This could discriminate against mobile operators in downstream retail markets that do not have a pan-European footprint and that lack the bargaining power. However, arbitrage by roaming brokers, new entry and wider geographical markets on the retail roaming level will work against this. Anti-discrimination rules contained in licenses, competition law or the GSM international roaming framework should not be applied in a way that impairs the competitive downward adjustment of wholesale roaming charges.

# 1 Introduction

Roaming can be defined as „facility, supported by commercial arrangements between operators and/or service providers, which enables a subscriber to use his/her radio telephone equipment on any other network which has entered into a roaming agreement in the same or another country for both outgoing and incoming calls."[1] When a subscriber uses his/her radio telephone equipment on a network in another country, the term „international roaming" is used.[2] International roaming allows subscribers of mobile networks to use their mobile phone and SIM (Subscriber Identity Module) card outside their home country and to make and receive calls abroad while still being billed by their home mobile operator or service provider ("one phone, one number, one bill"). This paper deals with international roaming on GSM networks, which is possible in 50 countries/areas in Europe and more than 170 countries/areas of the world; the focus of the paper is on international roaming in Western Europe.

The ability to make and receive roaming calls is of increasing importance to both residential and business customers. However, there are various complaints that prices of retail roaming are intransparent to consumers, rigid and at levels that are unrelated to the cost of carriage.[3] E.g., in November 1999 the International Telecommunications Users Group (INTUG) published a study showing that the difference in price between roamed and non-roamed mobile calls[4] for the same country-to-country pairs within the European Union can be as high as 500 percent, and a similar picture emerged in a follow-up study for 2000.[5] The Mobile Roaming Inquiry of the European Commission showed that, while prices of more competitive non-roamed mobile calls went down, prices of roamed calls often increased.[6] The UK regulator stated that the consumer does not appear to get a good deal on international roaming and that price competition on retail roaming calls is not evident.[7]

The primary focus of this paper is on wholesale roaming services, which are the prime driver of retail roaming prices.[8] The European Commission discovered that, over the period 1997 - 2000, mobile operators in many cases have substantially raised their

**1** European Commission (1994), p.225.
**2** "National roaming" describes the case where a subscriber roams on another network in the same country
**3** Retail roaming services are the services a home mobile operator offers its subscribers allowing them to use their subscription in other countries, by using the network of mobile operators in the visited countries.
**4** Viewed from a given network, "roamed calls" are those made by visiting subscribers, and "non-roamed calls" are those made by domestic subscribers. Technically, both calls are largely equivalent, with the prime exception that roamed calls require a signalling link to the subscriber's home network to check his/her status.
**5** INTUG (1999, 2001).
**6** European Commission (2000), p.3. See also Sauter (2001).
**7** Oftel (2000, 2001).
**8** Wholesale roaming services are the services a visited mobile operator offers to mobile operators licensed in other countries, allowing the subscribers of the latter to use the network of the former.

wholesale roaming rates, while, at the same time, prices of non-roamed (in particular, domestic non roamed) calls usually went down.[9]

The paper does not deal with actual market conduct of mobile operators and, in particular, cannot provide evidence on whether wholesale roaming charges are excessive or whether there was any collective price fixing. Rather, the thrust of this paper is on the structural conditions underlying wholesale roaming markets that have impaired incentives to engage in competitive price undercutting, and on developments that are likely to intensify competition on the wholesale roaming level in the future.

Chapter 2 gives a brief review of the basics of international roaming in the GSM world. Chapter 3 provides an analysis of supply-side conditions of wholesale roaming markets, with a focus on structural conditions that are likely to create oligopolistic interdependency between the leading providers of wholesale roaming services. Chapter 4 provides an analysis of demand side conditions that have impaired incentives to agree on lower wholesale roaming charges in exchange for preferred roaming status. The analysis of supply and demand side conditions also demonstrates the prospects for change and shows the potential for a competitive process of downward price adjustments. Chapter 5 concludes with a recommendation for application of non-discrimination rules when dealing with this evolving process.

## 2 Basics of international roaming

Wholesale roaming services are services a mobile operator of a given country ("visited network operator") offers to a mobile operator licensed in another country ("home network operator"), enabling the subscribers of the latter to use the network of the former. Wholesale roaming services include the provision of access to the visited network, and the provision of speech, data, fax and short message services (SMS) to the roaming subscriber. The provision of a roamed call to a foreign mobile operator's subscriber is technically similar to providing a non-roamed call to a domestic subscriber; it requires little extra functionality other than the signalling between the visited and home network. What is largely different, however, are underlying contracts and the marketing and billing relationships.

Wholesale roaming services are the major input for providing retail roaming services. Retail roaming services are the services a home mobile operator offers its subscribers allowing them to use their subscription in other countries, by using the network of mobile operators licensed in those countries. To ensure the best possible service to their customers, home network operators tend to maximise coverage by concluding international roaming agreements with (i) operators in a maximum number of countries and (ii) all mobile operators in a given country.

9 European Commission (2000), p. 18.

Wholesale roaming agreements are concluded on a commercial basis between individual licensed mobile operators that are members of the GSM Association, the industry body responsible for the development, deployment, evolution and promotion of the GSM standard. Wholesale roaming agreements are usually (but not necessarily) reciprocal, that is, both roaming partners reciprocally agree on the provision of wholesale roaming services. The GSM Memorandum of Understanding (MoU) provides the general basis for the establishment of international roaming, and the Standard International Roaming Agreement (STIRA), more explicitly, defines the principles of bilateral roaming agreements between GSM operators.[10]

The framework provided by the GSM MoU and STIRA does not deal with international roaming agreements between licensed GSM operators *and* organisations that are not licensed GSM operators, and, so far, there are no such international roaming agreements in place. Up to now, independant service providers buy international roaming services from licensed GSM operators in their home country within the framework of service provider agreements.[11] In order to enable a service provider's customers to roam abroad, GSM operators in the customer's home country must purchase wholesale roaming services from GSM operators in the visited countries and resell them to the service provider.

*Vertical relationships*

Figure 1 gives an illustration of the vertical relationships involved in international roaming. Assume that B1 denotes a customer's home network operator and A1 his/her visited network operator. Basically, there are two alternatives to distinguish:

(i) There is an international roaming agreement in place between mobile operators A1 and B1, and B1 purchases wholesale roaming services from A1. B1 may provide retail roaming services to its subscribers via an internal sales unit, or B1 may resell the wholesale roaming services purchased from A1 to a domestic service provider as part of a service provider agreement. The roaming agreement between mobile operators A1 and B1 may include an obligation on B1 to give A1 preferred roaming status. In that case, mobile operator B1 will initially programme its SIM cards in a way to direct its customers to A1's network once they enter A1's coverage area. Customers of B1 will automatically end up with the preferred roaming partner unless they manually select another network operator. Such a preferred roaming status is often agreed upon reciprocally, but unilateral obligations also exist, e.g., if incumbent operators require this as a condition for concluding a roaming agreement with a new entrant.

---

10 The GSM Association notified the STIRA in 1996, and the IOT in 1997. Both notifications received conditional exemptions from the cartel prohibition under Article 81 (3) of the EC Treaty.

11 The service provider agreements also encompass the wholesale provision of subscriptions and of national and international mobile calls to domestic subscribers (provision of "airtime").

(ii) Instead of concluding a roaming agreement with A1, mobile operator B1 may use a roaming broker. The roaming broker buys wholesale roaming services from mobile operator A1 (and other mobile operators) and resells them to B1. With the emergence of roaming brokers, mobile operators may buy roaming coverage for a large number of countries by contracting with a single entity. An example for a roaming broker is Comfone Ltd., whose roaming platform is based on Swisscom's roaming contracts.

**Figure 1:** Vertical relationships involved in international roaming

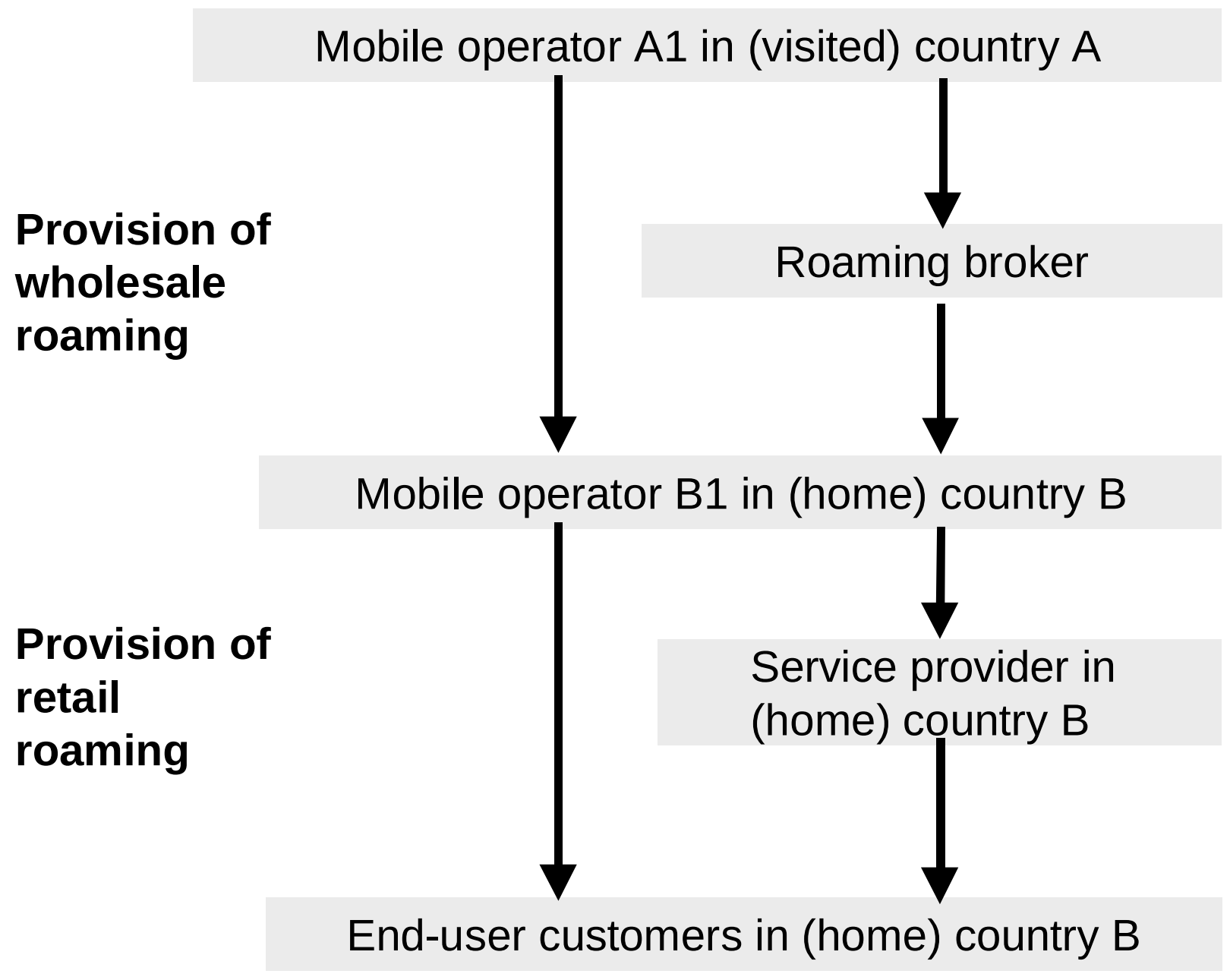

Source: WIK

*Pricing principles and billing relationships*

The general principles for setting wholesale roaming charges are defined by the GSM Association. Under the new regime in force since 1998/99, the wholesale roaming charge – called Inter-Operator Tariff (IOT) - is formally defined as "a tariff between mobile network operators, charged by the visited network operator to the home network operator for the use of the visited network" and is unrelated to the retail or wholesale prices of non-roamed calls.[12] The GSM framework requires mobile operators to apply

**12** Up to 1989/99, wholesale roaming charges for mobile originated calls were based on retail tariffs of non-roamed calls. This link to the visited mobile operator's prices of non-roamed calls no longer exists.

IOTs in a non-discriminatory uniform way to all foreign roaming partners. The framework does not prevent mobile operators from negotiating discounts, but such discounting appeared to be largely absent in the past.[13]

Under the new regime, both mobile originated and mobile terminated calls (outgoing and incoming roamed calls) can be charged for by the visited network operator. Figures 2 and 3 illustrate the inter-operator billing relationships:

(i) *Mobile originated roamed calls* (Figure 2): Assume a customer of mobile operator B1 roams on mobile operator A1's network and makes a call to his/her home country B, say to a customer of fixed network operator B2. Mobile network operator A1 usually hands over the call to a fixed network operator A2 in the visited country, which conveys the call to the customer's home country, where it is handed over to a fixed network operator B2, which terminates the call to the called subscriber. The inter-operator billing relationship are as follows: The fixed network operator B2 charges the fixed network operator A2 for terminating the call. The fixed network operator A2 charges the mobile network operator A1 for transiting the call and also recovers the termination charge paid to fixed network operator B2. The visited mobile operator A1 recovers its costs by charging its IOT for wholesale roaming calls to the customer's home mobile operator B1.

**Figure 2:** Billing relationships in case of mobile originated roamed calls

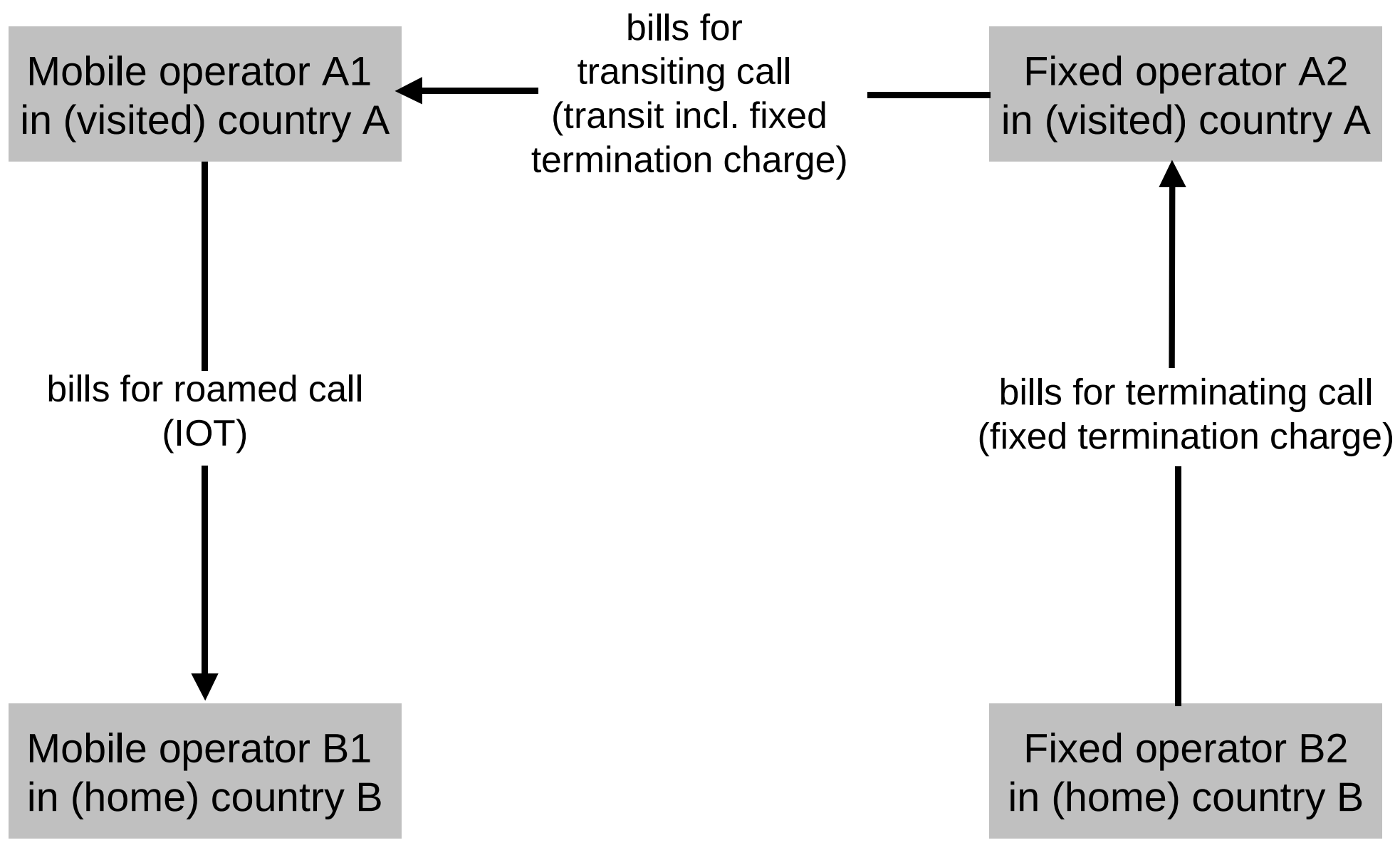

Source: WIK

13 See European Commission (2000), p. 8-9.

For mobile originated calls, IOT dimensions are usually destination (domestic or international), time of day (peak or off-peak), time unit (10 seconds/30 seconds/1 minute or other), type of terminating network, (fixed or mobile terminated) and/or may include a set-up fee for each call. For international destinations, operators usually use some form of zonal pricing, where a uniform IOT is set for a group of destinations.

(ii) *Mobile terminated roamed calls* (Figure 3)*:* Assume again that a customer of mobile operator B1 is roaming on mobile operator A1's network, but now receives a call from his/her home country made by a subscriber of fixed network operator B2. Fixed network operator B2 conveys the call to the visited country, where it is usually handed over to fixed network operator A2 that hands it over to the visited mobile operator A1 for termination. The terminating mobile operator A1 charges the fixed network operator A2 for terminating the call to the visiting customer. Fixed network operator A2 charges the fixed network operator B2 for transiting the call and also recovers the termination charge paid to mobile operator A1. Under the new IOT regime, the visited mobile operator A1 could also charge the customer's home mobile operator B1 for the call. Mobile terminated calls, however, continue to be zero priced, so mobile network operators have not exploited this possibility opened up by the new IOT regime.[14]

The GSM Association framework does not deal with pricing of roaming services on the retail level. Pricing principles are as follows:

- *Mobile originated roamed calls:* Traditionally, home network operators add a fixed percentage margin on top of the IOT, with the effect that retail roaming prices of a home network operator are a reflection of IOTs of visited network operators. The mark-up varies between 10 and 35% across countries, but is largely uniform within a given country. Only more recently, a number of mobile operators departed from this practice by offering single-rate retail tariffs for roaming in certain groups of countries, including discounted single-rate tariffs if subscribers roam on networks of affiliated firms.

- *Mobile terminated roamed calls:* If a customer is called from his/her home country while roaming on a foreign network, the home mobile operator charges its roaming customer the price of an international mobile call from the home country to the visited country.

**14** Under the former regime, a visited network operator could only charge roamers for mobile terminated calls if the visited network operator charged its own customers for such calls. In Western Europe this has not been the case. Under the new IOT regime, this restriction no longer applies. Mobile terminated call charges are now part of the wholesale inter-operator charging and no longer related to the way a visited network operator charges its own subscribers.

**Figure 3:** Billing relationships in case of mobile terminated roamed calls

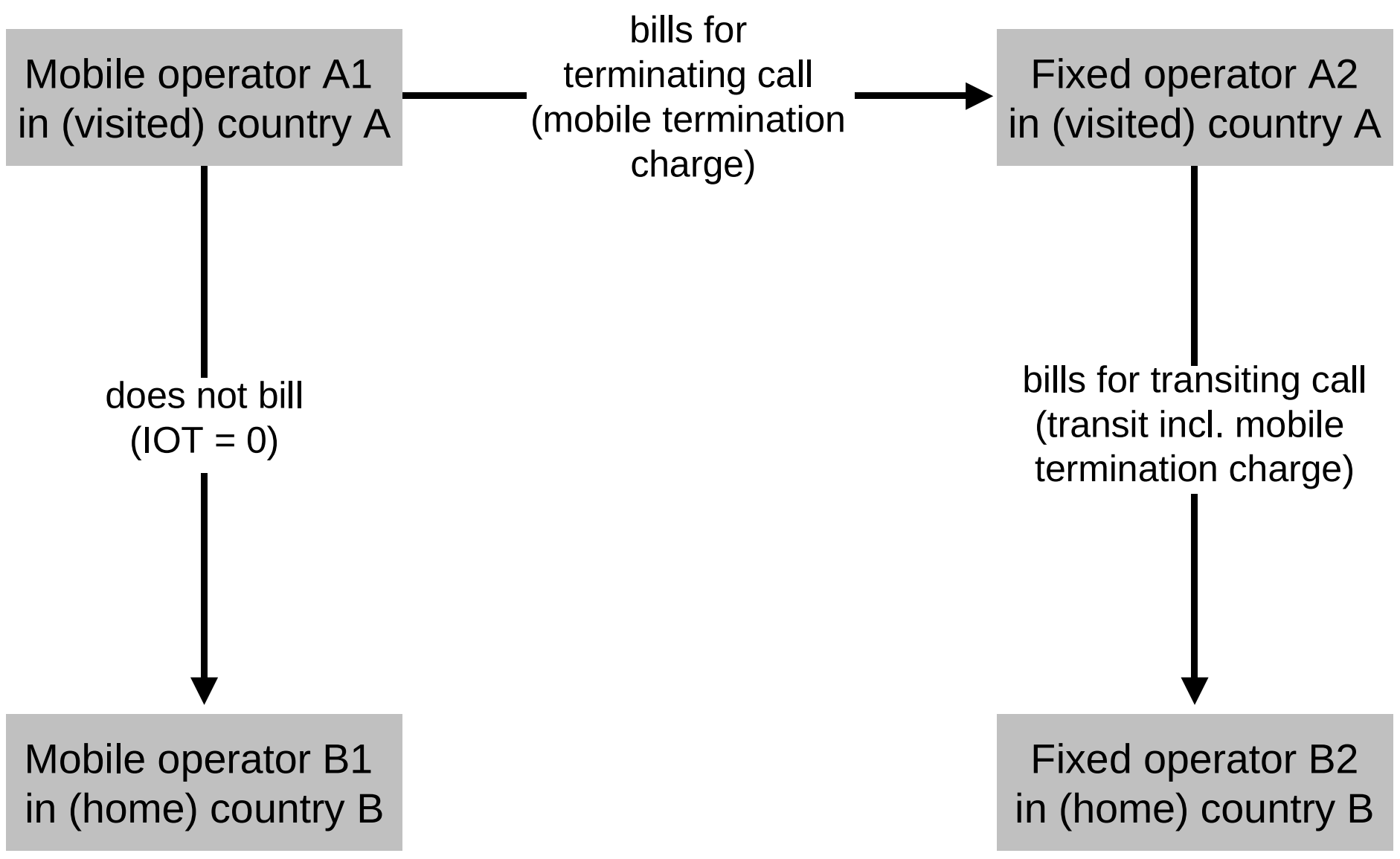

Source: WIK

## 3 Supply-side of wholesale roaming markets

Because, in Western Europe, mobile network licenses are accorded on a national basis, wholesale roaming markets are typically national. Each country constitutes a distinct national market for the provision of wholesale roaming services. The supply-side of wholesale roaming is characterised by a number of features which give rise to oligopolistic interdependency between the leading players and hardly provide incentives to engage in a process of competitive IOT undercutting. Those features include

(i) a small number of suppliers in each country and usually a high combined market share of the two leading operators,

(ii) barriers-to-entry and second-mover disadvantages that protect the leading operators,

(iii) imperfect substitutes to roaming relationships, and

(iv) a high transparency about IOTs that make competitive price undercuttings immediately visible.

This situation that has marked wholesale roaming markets since their creation may change in the future. Given the scarcity of spectrum, an increase in the number of GSM operators is not a prospect for most Western European countries. However, with second-mover disadvantages in wholesale roaming markets gradually disappearing, and price elasticity of individual firm's demand for wholesale roaming services increasing (the reasons are treated in chapter 4.2), new entrant GSM 1800 operators will get an increasing incentive to win market share by reducing their IOTs and/or offer discounts on IOTs. These developments are explored in more detail in this chaper.

## 3.1 Small number of suppliers and high market concentration

As Table 1 shows, the number of mobile operators with GSM licenses in Western European countries typically ranges from 3 to 4. The only countries with less than three licensed network operators are currently Luxembourg and Norway. Norway plans to issue new GSM licenses during autumn 2001, which will increase the number of GSM licenses to 3-4.

**Table 1:** Number of providers of GSM wholesale roaming services, Western Europe, August 2001

| | |
|---|---|
| Netherlands | 5 |
| Austria | 4 |
| Denmark | 4 |
| Finland | 4* |
| Germany | 4 |
| Italy | 4 |
| Liechtenstein | 4 |
| UK | 4 |
| Iceland | 3 (6**) |
| Greece | 3 (4**) |
| Belgium | 3 |
| France | 3 |
| Portugal | 3 |
| Spain | 3 |
| Sweden | 3 |
| Switzerland | 3 |
| Ireland | 3 |
| Norway | 2 (3-4***) |
| Luxembourg | 2 |

* GSM operators with local licenses not included.

** Number in brackets also includes GSM mobile operators, which have already received a license, but did not yet commence service.

*** Number in brackets also includes new GSM mobile operators, which are likely to be licensed in autumn 2001.

Source: Mobile Communications, GSM Association, WIK

In each country, the provision of wholesale roaming is generally highly concentrated. The Mobile Roaming Inquiry of the European Commission showed that, in a large number of countries in Western Europe (Austria, Belgium, Denmark, France, Germany, Iceland, Italy, the Netherlands, Norway, Spain, and the United Kingdom), the leading operator has an estimated market share of over 50%. The two largest mobile operators' combined share of the national wholesale roaming market is typically above 90%.[15] The leading mobile operators are typically the incumbent GSM 900 operators licensed first, and those with small or negligible market shares are the new entrant GSM 1800 operators licensed at a later stage.[16]

## 3.2 Spectrum scarcity and second-mover disadvantages

*New entry*

The mobile sector as a whole is characterised by barriers-to-entry of which the scarcity of spectrum for GSM mobile telecommunications services is the most prominent. The lack of spectrum limits the number of GSM licenses and, in most Western European countries, makes it unlikely that additional GSM licenses will be made available in the future.

Hopes have been placed on mobile virtual network operators (MVNOs) and new entrant mobile network operators with UMTS licenses. However, since both of them do not dispose of GSM spectrum, they will have to use capacity on an existing GSM operator's radio access network.

(i) MVNOs have some physical network infrastructure comprising as a minimum a switching centre, a home location register (HLR) and authentication centre. They have their own unique mobile network code and issue their own branded SIM cards. MVNOs take over their customers's calls and route them to the final destination on the basis of interconnection contracts with other fixed and/or mobile network operators. Hence, MVNOs are different to service providers providers which deliver services entirely over other mobile operators' networks. Various service providers claim to be MVNOs, but since they do not meet the minimum requirements described above, a term such as "enhanced service providers" would be more appropriate for them. An MVNO in the strict sense appears to exist in Denmark (Tele2). In Germany, UMTS newcomers Mobilcom and 3GMobile will get MVNO status for GSM.

---

15 European Commission (2000), p. 23.

16 It should be noted that, in various countries, incumbent GSM 900 operators now have also been allocated additional spectrum in the 1800 MHz range, and likewise GSM 1800 new entrant operators have been allocated spectrum in the 900 MHz range. We will, however, continue to use the traditional terms of "GSM 900 operator" usually denoting an imcumbent, and "GSM 1800 operator" usually denoting a new entrant licensed at a later stage.

(ii) In some countries, the spectrum made available for UMTS allowed to grant UMTS licenses in a number that exceeds the number of existing GSM licenses (e.g., in Germany, Austria, Italy, Norway, Portugal, Spain, Sweden, Switzerland, and UK). This leads to entry of new mobile operators. Also, in a few countries, newcomers outbid GSM operators in UMTS auctions (e.g., in Italy and Denmark). As a result of regulatory obligations or commercial agreements, newcomers will be able to use capacity on existing GSM networks during roll-out of their UMTS networks. This is usually called 3G-2G national roaming.

However, neither MVNOs nor UMTS newcomers will be able to provide wholesale roaming services to international roaming partners. Both of them do not dispose of GSM spectrum, and neither MVNO arrangements nor national roaming agreements will allow the reselling of wholesale capacity purchased on existing GSM networks to foreign roaming partners. The number of providers of GSM wholesale roaming services, therefore, will not increase by those developments.

*Second mover disadvantages*

The practice of most Western European countries of sequentially issuing GSM licenses (illustrated in Table 2) created various second-mover disadvantages in the mobile sector. The persistently high combined market share of the two leading mobile operators (usually GSM 900 operators) in wholesale roaming markets is a reflection of this. New entrant GSM 1800 mobile operators licensed at a later stage had to face two important disadvantages when trying to generate wholesale roaming traffic on their networks:

(i) Unavailability of dual band GSM 900 and 1800 handsets disadvantaged new entrant GSM 1800 operators when selling wholesale roaming to incumbent GSM 900 operators and seeking preferred roaming status from them. Initially, the majority of subscribers of GSM 900 networks could not roam on GSM 1800 networks due to their handsets that were only 900 MHz compatible. This was also a reason, why GSM 900 operators, initially, were less interested in concluding roaming agreements with GSM 1800 operators. However, as 1800 MHz networks are rolled out, and as GSM 900 operators are also awarded 1800 MHz frequencies, dual band handsets are now becoming widely deployed. GSM 900 operators now perceive a benefit from offering roaming on 1800 MHz networks in other countries to their subscribers. In fact, where GSM 1800 operators have rolled out their networks, they may provide higher signal strength, better voice quality and higher availability rates than GSM 900 operators. As a result, any disadvantage for new entrant GSM 1800 operators related to availability of dual band handsets is likely to disappear. The ability of new entrant GSM 1800 operators to compete for wholesale roaming market share with incumbent GSM 900 operators will increase due to the greater penetration of dual-band handsets.

**Table 2:** Date of launch of GSM networks in Western Europe, August 2001

| | Rank in market entry | | | | | |
|---|---|---|---|---|---|---|
| | 1st | 2nd | 3rd | 4th | 5th | 6th |
| **Austria** | Mobilkom (12/93) | Max.mobil (10/96) | One (10/98) | Tele.ring (5/00) | - | - |
| **Belgium** | Belgacom (1/94) | Mobistar (8/96) | KPN Orange (3/99) | - | - | - |
| **Denmark** | TDC (7/92) | Sonofon (7/92) | Telia (1/98) | Mobilix (3/98) | - | - |
| **Finland** | Radiolinja (12/91) | Sonera (7/92) | Telia (3/98) | Suomen 2G (2/01) | - | - |
| **France** | Orange (7/92) | SFR (12/92) | Bouygues (5/96) | - | - | - |
| **Germany** | D2-Vodafone (6/92) | T-Mobil (7/92) | E-Plus (5/94) | VIAG Interk. (10/98) | - | - |
| **Greece** | Panafon (7/93) | Stet Hellas (7/93) | Cosmote (3/98) | Info-Quest (na) | - | - |
| **Iceland** | PTT (12/94) | Tal (5/98) | Íslandssími (3/01) | IMC Ísland (planned 1/02) | Halló!-Frjáls (planned) | Lína.Net (planned) |
| **Ireland** | Eircell (7/93) | Esat Digifone (3/9) | Meteor (02/01) | - | - | - |
| **Italy** | TIM (4/95) | Omnitel Pronto (10/95) | Wind (3/99) | Blu (5/00) | - | - |
| **Liechten-stein** | Telecom FL (na) | Tele2 (3/00) | VIAG Europl. (8/00) | Mobilkom (9/00) | - | - |
| **Luxem-bourg** | P+T (7/93) | Millicom (5/98) | - | - | - | - |
| **Nether-lands** | KPN Mobile (7/94) | Libertel-Vodafone (9/95) | Telfort (10/98) | Dutchtone (1/99) | Ben (2/99) | - |
| **Norway** | Telenor (5/93) | NetCom (9/93) | - | - | - | - |
| **Portugal** | TMN (10/92) | Telecel-Vodafone (10/92) | Optimus (9/98) | - | - | - |
| **Spain** | Telefónica Móviles (7/95) | Airtel (10/95) | Amena (1/99) | - | - | - |
| **Sweden** | Tele2 (9/92) | Europolitan (9/92) | Telia Mobile (11/92) | - | - | - |
| **Switzer-land** | Swisscom Mob (3/93) | Sunrise (12/98) | Orange Switzerl. (6/99) | - | - | - |
| **UK** | Vodafone (7/92) | One-2-One (9/93) | BT Cellnet * (1/94) | Orange UK (4/94) | - | - |

* Although BT Cellnet was only third in launching its GSM network, it must be regarded as an incumbent since it had a license for NMT (analogue mobile telecommunications) networks and was well established as a mobile operator before One-2-One and Orange entered the market. Vodafone, which was first in launching a GSM network also had a NMT license.

Source: Mobile Communications, GSM Association, WIK

(ii) Lower coverage rates during network roll-out were a second major handicap for new entrant GSM 1800 operators, when seeking preferred roaming status from roaming partners and/or attracting foreign subscribers to their networks. In a number of countries (e.g., Denmark, Iceland, Italy, Spain, Norway), GSM 900 operators are required to provide national roaming to GSM 1800 operators during a transitional period, and in others (e.g., Germany) network operators provide national roaming for commercial reasons. However, buyers of national roaming services usually are not allowed to resell these services to foreign network operators. National roaming, whether mandated or commercial, allows a network operator during network roll-out to increase its coverage *vis-à-vis* domestic subscribers, but *not* in relation to roaming subscribes of foreign network operators. Despite national roaming, new entrants in the wholesale (international) roaming market are severely disadvantaged. Without nationwide coverage, it is more difficult to get preferred roaming status from foreign roaming partners and/or generate roaming traffic on one's network. Again, this problem will disappear in the longer term: GSM 1800 new entrants in many countries will have soon rolled out their networks to nationwide coverage and, hence, will be able to provide the same quality of wholesale roaming services to foreign roaming partners as incumbent GSM 900 operators.

## 3.3 Imperfect substitutes to wholesale roaming

Apart from call-back applications, there are no alternatives to roaming relationships. With call-back applications, home network operators can enable their subscribers to make mobile calls in other countries outside traditional roaming agreements. Call-back applications can substitute outbound roaming calls by calls in a reverse direction. Calls in a reverse direction are set up by the home operator itself and terminated using interconnection agreements. Call-back applications of European mobile operators are based on USSD (Unstructured Supplementary Service Data). USSD is a means of transmitting information over a GSM network which has some similarities with SMS, since both use the GSM network's signalling path. Call-back applications are already marketed in two variants:

(i) as a pre-pay service addressed at international customers with an additional SIM card (e.g., Swisscom provides such a service under the "EasyRoam" brand for use outside of Switzerland),

(ii) as a pre-pay service addressed at domestic residential customers for use with their regular SIM card (e.g., T-Mobil).

Call-back solutions, however, cannot be regarded as a full substitute to roaming. First, from a subscriber's perspective, alternative (i) may be regarded as inferior, because it requires an additional SIM and the allocation of a new mobile number. This, however, might no longer be an obstacle in the longer term, following the introduction of multi-SIM handsets. Second, and more important, call-back solutions can only substitute mobile originated calls to the subscriber's home country. Domestic calls within the visited country or mobile terminated calls (incoming calls) necessitate a traditional roaming arrangement between visited and home network operator. Hence, from a home network operator's perspective, call-back solutions are only a partial and imperfect substitute for traditional wholesale roaming arrangements.

### 3.4 Transparency of competitors' IOTs

Wholesale roaming markets are characterised by a high degree of transparency about IOTs. Mobile operators notify any change of IOTs through the GSM InfoCentre to roaming partners. The IOTs of a particular mobile operator listed on the InfoCentre are electronically accessible to all GSM MoU members, with the exclusion of competitors in the domestic wholesale roaming market. Competitors, however, may usually find out about the IOTs offered by competing mobile operators through affiliated operators in other countries. There is also a straightforward way to calculate competitors' IOTs, simply by deducting the retail margin of 10 to 35% (which is usually publicly known) from published retail roaming prices. Information about IOTs, if available routinely in the market, makes IOT reductions immediately visible to competitors and reduces incentives to competitive price undercutting.

A different situation exists with regard to individual discounts on IOTs, which must not be revealed on the InfoCentre. Such discounting has been largely absent in the past, but if used more intensely in the future (see the following chapter 4.2), this would reduce price transparency on the supply-side and would provide incentives for more price competition.

## 4 Demand-side of wholesale roaming markets

Oligopolistic interdependence is not the sole possible reason for the price rigidity that characterised wholesale roaming markets in the past. It also does not explain the fact that new entrant GSM 1800 operators with negligible market shares, in many instances, charge IOTs above IOTs of incumbent GSM 900 operators. The reason is to be found on the demand-side of wholesale roaming. As is discussed on more detail below, foreign roaming partners

(i) face little competition in retail roaming markets at home, and, therefore, little pressure to seek more favourable IOTs in order to cut costs,

(ii) have insufficient control over their subscribers' network selection in the visited country, which makes demand for roaming on a given visited network largely insensitive to price changes, gives rise to demand externalities and reduces incentives to offer preferred roaming status and/or traffic growth in exchange for discounts on IOTs.

This situation is likely to change for two reasons described in more detail below. First, retail roaming markets are likely to become more competitive with new entry of mobile operators, and business markets growing into a pan-European dimension. Second, and even more important, the introduction of over-the-air programming of SIM cards will enable home mobile operators to direct their customers to the visited networks with the lowest IOTs. This will remove demand externalities and create incentives to agree on discounts on IOTs in exchange for preferred roaming status and traffic growth. On the supply side of wholesale roaming markets, the incentive to agree on such discount schemes should be particularly pronounced for new entrant GSM 1800 mobile networks operators with so far negligible market shares.

## 4.1 Lack of competitive pressure in downstream retail roaming markets

So far, structural conditions of retail roaming markets are not conducive to price competition. A high combined market share of the two leading players, together with pronounced second-mover disadvantages of new entrant GSM 1800 operators and other providers of retail roaming services explains the lack of competition.

*High concentration*

Compared with wholesale roaming services, the number of suppliers of retail roaming services in a given country is usually higher, and concentration rates are lower. This is due to the existence of independent service providers. The combined share of the two leading mobile operators of retail roaming revenue is usually still over 60%, compared with over 90% for wholesale roaming.[17]

*Barriers-to-entry and second-mover disadvantages*

Because of spectrum scarcity, the number of GSM licences is limited, and in most countries additional GSM licenses are unlikely to be issued. Nevertheless, as discussed above, new types of providers of GSM services can enter the market:

(i) MVNOs have appeared in Denmark (Tele2) and will soon appear soon in Germany (Mobilcom and 3GMobile).

(ii) Allocation of UMTS licenses will in some countries increase the number of mobile operators by one or two.[18]

---

17 European Commission (2000), p. 17.

18 In Germany, this is not regarded as „national roaming“, although it may be technically identical. In regulatory terms, roaming as defined in Germany is restricted to the same market and aims to extend

Both MVNOs and UMTS newcomers will be able to conclude GSM (and UMTS) international roaming agreements in order to provide retail roaming services to their customers.

New entrants face substantial second-mover disadvantages as can be seen in the case of existing GSM 1800 operators. They are far less successful in extracting retail roaming revenue from their customer base due to the preponderance of residential and prepaid customers (which reflects later market entry). Residential contract customers typically generate less roaming revenue per user than business customers. Also, in the past, roaming services could not be made available to pre-pay customers. This only changes now, with the introduciton of Intelligent Network functionality (CAMEL[19]) that allows mobile operators to increase their control of roaming activity and to promote offerings of pre-pay roaming.

Other ways of market entry are possible on the retail level: Firms may enter as service providers or indirect access providers. The competitive impact, however, is likely to be very limited:

(i) Independent service providers contribute little in intensifying price competition for retail roaming services. Because service providers cannot directly conclude roaming agreements with mobile operators in other countries or with a roaming broker, they simply resell roaming services purchased by the home network operator they have a service provision agreement with. When customers of service providers roam abroad, all calls are billed to the relevant home network operator by the visited network operator. The home network operator adds a handling charge before billing the roamed calls to the domestic service provider. Home network operator and service provider share the handling charge.

Service providers would have a greater impact on competition if they could conclude roaming agreements. Since they would be too small to conclude multiple roaming agreements with a large number of mobile network operators, they would have to have access to roaming brokers (which so far is not the case). Roaming brokers would have a central role in providing wholesale roaming access to service providers.

(ii) Some countries have introduced mobile carrier selection (e.g., Finland, Denmark, Germany, Spain, UK and Norway), but there does not seem to have been market entry by indirect access operators, except in Finland. If mobile carrier selection was available, subscribers could use their existing connection to a domestic mobile network operator to route calls to a selected operator (called “indirect access operator”). Customers would continue to have a

---

the availability of an operator’s network for his customers. An UMTS license holder, however, cannot sign a GSM national roaming agreement with a GSM license holder extending the availability of his network (roaming), quite simply because he does not have a GSM mobile network. See Kurth (2001).

19 Customised Applications for Mobile Network Enhanced Logic.

subscription with their access operator, but would be able to make calls with the indirect access operator. A customer could use indirect access operators on a call-by-call basis by using a prefix before the dialled number (or on a pre-selection basis, if available in the country).

As far as retail roaming calls are concerned, indirect access will be of a limited impact. As usually understood, carrier selection allows *domestic* customers to use an existing connection with a *domestic* mobile network operator to route domestic or international calls through another operator (indirect access operator). A *roaming* customer, however, would have to use *roaming* access in a visited network to route calls through a third operator. So far, the technical, commercial and regulatory basis appears to be uncertain. But even though carrier selection and indirect access may not be applicable for roaming calls, it could still have an indirect effect through lowering prices for domestic and international mobile calls. The alternatives to roaming calls discussed below – calling cards and pre-paid cards – could become more attractive substitutes and impose pressure on retail roaming prices.

*Imperfect substitutes to retail roaming*

From an end-customer's perspective, retail roaming services are only partially and imperfectly substitutable by call-back services or pre-pay services bought in the visited country.

(i) International call-back services are an alternative to outbound roaming calls. Some are marketed on a European-wide basis to business customers as an additional service to be used beside the customer's regular subscription (with a separate SIM), some are marketed to domestic residential customers as part of regular pre-pay packages. The introduction of multi-SIM handsets could facilitate the use of alternative SIM cards and make call-back services a good substitute for outbound international roaming calls.

(ii) Another alternative for customers when travelling abroad is to purchase a pre-pay card from a mobile operator licensed in the visited country and use it in the GSM handset instead of the SIM card of the home network operator (this is sometimes called "plastic roaming"). The price of a pre-pay international call may in some cases be higher than the rate per minute of a roamed international call. However, the price of a pre-pay domestic call appears to be usually lower than the rate per minute of a roamed domestic call. Also, in contrast to incoming roamed calls, no charge is to be paid for incoming calls in case of a pre-pay card. Hence, the savings on domestic calls and incoming calls can justify purchasing a pre-pay card. The introduction of multi-SIM handsets will also facilitate "plastic roaming".

## 4.2 Customer ignorance, insufficient control over network selection, and demand externalities

Mobile operators usually offer their subscribers the choice between roaming on several networks in other countries. For pricing retail roaming services, two approaches are applied:

(i) The traditional way is to charge subscribers for roaming depending on which network subscribers actually roam onto. In this case, mobile operators put a fixed percentage margin on top of the IOT paid to the visited network operator.

(ii) Alternatively, some mobile operators offer single-rate tariffs for roaming in a particular group of countries, with discounted single-rate tariffs if calls are made using the network of an affiliated operator.

The first approach necessarily entails a low degree of transparency on the subscribers' side. Where retail roaming prices are set by marking up wholesale roaming charges, there are thousands of possible variations of international roaming charges in Europe alone. In the past, mobile operators made little effort to inform their customers about relative prices of roaming on alternative networks in a visited country. The result is that customers are usually ignorant about the prices applied when roaming abroad. In addition, many subscribers are not familiar with manual network selection that would allow them to switch to the cheapest network.

When a customer enters a country, he/she usually leaves it to the handset to automatically choose a network. SIM cards contain a pre-programmed preferred list of networks. The handset looks at the preferred list and searches for the first network on the list. If it finds the network's signal, it will log on this network. If not, it will turn to another one depending on signal strength. Although mobile operators can direct their subscribers to preferred networks through initial programming of the SIM card, they are unable to make any subsequent modifications. The preferred list in the SIM card is not updated to reflect changes in IOTs. Customer ignorance about relative retail roaming prices and manual selection of networks make the choice of the visited network largely dependent on the original programming of the SIM card.

Of cause, customers will usually control their roaming bill *ex post* and may have a rough perception of the average price per minute of roaming in a particular country. This will have an influence on the number of roaming calls in a visited country and the volume of roaming minutes, but not on the network selected.

Customer ignorance and operators' lack of control over network selection gives rise to an externality. To demonstrate this, assume that

(i) a visited mobile operator A1 decreases its IOT below the level of competing mobile operators in country A;

(ii) the home mobile operators in country B reduce the retail price for roaming on network A1 by the same percentage rate, but are unable to redirect their subscribers to mobile operator A1, and

(iii) subscribers of home mobile operators in country B remain ignorant about the change in relative retail roaming prices, but take note of a decrease in the average price per minute of a roaming call in the visited country (as a result of inspecting their monthly bill).

A decrease in A1's IOT, with a corresponding decrease in the retail prices of roaming on A1's network, will lead to a decrease of the perceived average price of roaming in country A. This will increase the total quantity of minutes roamed on networks in country A, but shares of visited network operators in country A will remain roughly the same. The decrease in A1's IOTs will benefit both A1 and its competitors, depending on their market shares.

This externality provides disincentives for mobile operators to lower IOTs. The smaller a mobile operator (in terms of its share of the national wholesale roaming market), the less likely it is to internalise the demand increasing effect of a decrease of its IOT. As a result, smaller operators, in particular, have an incentive to charge high IOTs, since a reduction in IOTs would hardly affect the volume of roaming minutes they can generate on their networks.[20] Whereas a large mobile operator must account for the impact of its IOT on its roaming volume, a small operator faces a very inelastic demand for roaming and thus can impose higher mark-ups above the marginal costs of providing roaming calls. A mobile operator may have a small market share in its wholesale roaming market. Yet under the assumption that customers base their roaming decisions only on perceived average retail roaming prices, a small visited mobile operator has market power in its wholesale roaming market. Indeed, a mobile operator can refrain from following other competitors in reducing IOTs and hardly suffer a reduction in its share of the wholesale roaming market.

GSM Europe, the European Interest Group of the GSM Association, recently adopted a "Code of Conduct for Information on International Retail Roaming Prices" to promote existing best practices, such as providing information over the customer service number or the Internet site of the home operator, and to encourage other options such as information via SMS, Fax on demand, e-mail, WAP, roaming guides, information material at points of border entry/exit, leaflets accompanying the bill, and information via retail outlets. Such measures can increase users' awareness of international retail

**20** A conceptually similar problem occurs in fixed-to-mobile calls, if the calling party is ignorant about the terminating operator and, hence, about the price of the call. As a result of this ignorance, the end-user relies on an estimated average price of fixed-to-mobile calls to all mobile operators. Consumer ignorance implies that mobile operators can increase the charge for mobile termination without feeling the full demand reducing effect of the increase. This has been explored, in particular, by the Australian ACCC (2001). For a general exposition of the theory behind, see e.g. Laffont and Tirole (2000), p. 184 – 187.

roaming issues. However, given the complexity of the retail roaming tariff structures involved, it is debatable whether existing externalities can be remedied by more information alone.

Recently, a number of operators have started to offer averaged retail roaming prices, which give customers a single rate for roaming in a range countries, e.g., Vodafone, Orange, T-Mobile, BT Cellnet, KPN Mobile, and TIM have introduced such tariffs. Single-rate tariffs have the advantage of providing greater transparency of tariffs and allow users to better compare offerings between mobile operators when choosing a subscription. However, they do not guarantee that customers roam on the least-cost network. In fact, they change nothing as far as the arbitrary nature of network selection is concerned. Single-rate tariffs will not provide incentives for IOT reductions or discounting as long as home network operators cannot influence network selection of their subscribers.

The market externality can be removed by the introduction of SIM Application Toolkit in combination with over-the-air-programming of the SIM card. The SIM Application Toolkit extends the role of the SIM card, and makes it a key interface between the mobile terminal and the network. Using the SIM Toolkit, the SIM can be programmed over the air to modify the list of preferred networks. Customers can then be redirected to the network which has lowered its IOTs, or they can be moved between networks to benefit from peak/off-peak differentials.[21] The introduction of SIM Application Toolkit and SIM over-the-air programming will give home mobile operators control over network selection of their subscribers. Visited network operators will have an incentive to offer discounts in return for being granted preferred roaming status. Such discounting can be based on the volume of revenue or minutes, or the growth in the volume of revenue or minutes. This can benefit new entrant operators which will find it more beneficial then now to offer discounts and/or undercut incumbents' IOTs. On the demand side, the driving force will probably be integrated pan-European mobile operators, which can offer the combined roaming volume and growth of affiliated operators active in several countries. Table 3 illustrates the potential of Vodafone and other pan-European operators in combining demand of affiliated companies for wholesale roaming.

**21** AT&T Wireless has developed an Intelligent Roaming Database which ranks the list of mobile operators by priority, and which is updated and downloaded over the air into each digital multi-network phone every month.

**Table 3:** Affiliated firms of pan-European mobile operators, Western Europe, September 2001 (percentage direct or indirect ownership)

| | Vodafone | Orange | T-Mobile | mmO2 | TIM | KPN | Telia |
|---|---|---|---|---|---|---|---|
| **Austria** | - | Connect Austria (17,5%) | Max.mobil (100%) | - | Mobilkom (25%) | - | - |
| **Belgium** | Belgacom (25%) | Mobistar (50,7%) | - | - | - | KPN Orange (100%) | - |
| **Denmark** | - | Mobilix (53,6%) | - | - | - | - | Telia DK (100%) |
| **Finland** | - | - | - | - | - | - | Telia Finland (100%) |
| **France** | SFR (31,9%) | Orange France (100%) | | SFR (20,8%) | Bouygues (10,7%) | - | - |
| **Germany** | D2-Vodafone (99,2%) | - | T-Mobil (100%) | VIAG Interk (100%) | - | E-Plus (77,5%) | - |
| **Greece** | Panafon (52,8%) | - | - | - | Stet Hellas (52,29%) | - | - |
| **Ireland** | Eircell (100%) | - | - | Esat Digif. (100%) | - | | - |
| **Italy** | Omnitel Pronto (76,1%) | Wind (43,4% ) | - | Blu (20 %) | TIM | - | - |
| **Nether-lands** | Libertel-Vodafone (70%) | Dutchtone (100% ) | Ben (49,9%) | Telfort (100%) | - | KPN Mobile Netherlands (100%) | - |
| **Norway** | - | - | - | - | - | - | NetCom (100%) |
| **Portugal** | Telecel-Vodafone (50,9%) | Optimus (20% ) | - | - | - | - | - |
| **Spain** | Airtel (91,6%) | - | - | - | Amena (26,4%) | - | -- |
| **Sweden** | Europolitan (71,1%) | - | - | - | - | - | Telia Mobile (100%) |
| **Switzer-land** | Swisscom Mobile (25%) | Orange Comms. Switzerl. (85%) | - | - | - | - | - |
| **UK** | Vodafone UK (100%) | Orange UK (100%) | One-2-One (100%) | BT Cellnet (100%) | - | - | - |

Source: WIK

# 5 Conclusions and implications for application of non-discrimination rules

The developments described above are likely to modify the structural conditions of wholesale roaming markets and initiate a process of competitive discounting and IOT reductions.

(i) With the introduction of SIM application toolkit and SIM over-the-air programming, home mobile operators will be able to direct customers to networks with the lowest IOTs. This will impose incentives on visited mobile operators to offer discounts on IOTs in exchange for preferred roaming status and traffic growth. The demand externality that currently exists will disappear. The incentive to lower IOTs and offer discounts will be particularly large for smaller new entrant GSM 1800 operators.

(ii) As dual mode handsets become ubiquitous and as existing GSM 1800 operators reach nationwide coverage, second-mover disadvantages for GSM 1800 operators will disappear. When seeking preferred roaming status from roaming partners and/or attracting foreign subscribers to their networks, they will be able to compete with similar coverage rates with incumbent GSM 900 operators. Given the small shares of the wholesale roaming market that GSM 1800 operators currently dispose of, they should have a particular incentive to win market share by offering discounts on IOTs and/or lowering IOTs.

This is likely to initiate a process of IOT undercutting, where the leading GSM 900 operators will have to follow.

On the demand side, it will be the GSM 900 operators that will put pressure on IOTs and ask for discounts given their larger retail roaming volumes. It may be that pan-European mobile operators combine roaming volumes of affiliated operators in various countries to get higher discounts. This could discriminate against home mobile operators that do not have a pan-European footprint. However, there are two factors which work against discrimination:

(i) In the longer term, price discrimination assumes that mobile operators have the ability to segment foreign roaming partners and, more importantly, prevent arbitrage occurring between them. Roaming brokers, even though they are unlikely to accumulate as much roaming traffic as the larger pan-European operators could be an important force that works against discrimination.

(ii) Competition in the provision of retail roaming is likely to increase with market entry of new mobile operators (MVNOs, UMTS newcomers that will also offer GSM roaming to their customers), and as a result of the creation of a pan-European market for roaming services for internationally mobile business customers. If retail roaming became more competitive, the bargaining power of pan-European operators would not create competition policy problems. This, however, would have to be studied more thoroughly in the light of future developments.

The GSM framework requires that mobile operators apply their IOTs in a non-discriminatory way to all foreign roaming partners. This echoes non-discrimination provisions in competition and telecommunication laws and licensing conditions of mobile operators. However, a restrictive interpretation of these provisions may prevent roaming partners, and in particular new entrant GSM 1800 operators, from offering discounts in exchange for preferred roaming status and/or for additional traffic volume. Also, non-discrimination rules should not prevent mobile operators with large numbers of roaming subscribers to put pressure on wholesale roaming charges by asking for discounts, provided retail roaming markets become more competitive.[22] As long as retail markets are not sufficiently competitive, it is important that discounts offered to affiliates of pan-European operators on the basis of volume and/or traffic growth are made available to all foreign roaming partners including roaming brokers.[23]

---

**22** The European Commission (2000), p. 24, also acknowledges that the principle of non-discrimination applied by licensing, competition and regulatory rules may act as a disincentive to price wholesale roaming services more competitively.

**23** The danger of discrimination and foreclosure led the European Commission to impose a non discrimination obligation on Vodafone as a result of its merger with Mannesmann. Pursuant to the Vodafone AirTouch/Mannesmann decision, Vodafone has to provide third parties non-discriminatory access to discounted IOTs up to April 2003. Due to these undertakings, Vodafone and its subsidiaries are obliged to make discounted IOTs available to other mobile operators. See European Commission (2001).